\documentclass[twocolumn,superscriptaddress,nobibnotes,nofootinbib,floatfix,aps,prl,citeautoscript,reprint]{revtex4-2}

\usepackage{graphicx}
\usepackage{amsmath,bm}
\usepackage{amsfonts}
\usepackage{dcolumn}
\usepackage{physics}
\usepackage{float}
\usepackage[colorlinks,linkcolor=blue,citecolor=blue,urlcolor=blue]{hyperref}

\newcommand{\portland}{Department of Mechanical and Materials Engineering, Portland State University, Portland, OR 97201, USA}

\begin{document}

\title{Dynamical Phonon Anharmonicity Drives Charge-Density-Wave Softening and Electronic Quasiparticle Decoherence in Monolayer 1H-NbSe$_2$}

\author{Yi Xia}
\email{yimaverickxia@gmail.com; yxia@pdx.edu}
\affiliation{\portland}

\date{\today}

\begin{abstract}
Charge-density-wave (CDW) instabilities in two-dimensional materials are governed by strongly anharmonic soft phonons, whose dynamical behavior remains poorly understood beyond the statically renormalized vibrations. Here, using monolayer 1H-NbSe$_2$, we investigate how frequency-dependent three- and four-phonon interactions reshape phonon spectral functions and influence electronic excitations. We find that three-phonon interactions induce substantial dynamical softening of CDW modes, increasing the estimated transition temperature relative to static self-consistent phonon theory. Four-phonon interactions further promote such dynamical softening and enhance spectral broadening, producing non-Lorentzian lineshapes that challenge the quasiparticle picture. Incorporating dynamically renormalized phonon frequencies and electron–phonon coupling into Fan-Migdal electron self-energy calculations reveals enhanced electronic kinks, satellite features, and momentum-dependent quasiparticle decoherence, driven by the strong momentum dependence of the electron-phonon coupling matrix elements. These findings establish dynamical anharmonicity as a critical link between soft-mode lattice fluctuations and electronic spectral signatures, providing insights into coupled lattice-electron dynamics in two-dimensional quantum materials.
\end{abstract}

\maketitle

\section{Introduction}

Charge-density-wave (CDW) instability, usually found in low-dimensional metals, is a fundamental many-body state, in which a spontaneous charge modulation couples to a periodic lattice distortion~\cite{Wilson1975CDW,Rossnagel2011CDW}. In layered transition metal dichalcogenides (TMDs), CDW phases often coexist or compete with superconductivity and other correlated states, providing a versatile platform for exploring intertwined orders and for tuning electronic structure in two-dimensional devices~\cite{Wilson1975CDW,Manzeli2017TMD,Xi2015monolayerCDW,Ugeda2016NbSe2}. Among TMDs, NbSe$_2$ is of particular interest, as bulk 2H-NbSe$_2$ undergoes an incommensurate CDW transition at $T_{\rm CDW}\approx33$~K and exhibits superconductivity below $T_c\approx7.2$~K~\cite{Wilson1975CDW,Moncton1975NbSe2,bevolo1974specific,alshemi2025two}. In the monolayer limit, Raman and transport experiments report a strongly enhanced $T_{\mathrm{CDW}}\approx145$~K together with a suppressed superconducting $T_c$~\cite{Xi2015monolayerCDW}, while scanning tunneling microscopy confirms that a $3\times3$ CDW order survives down to a single layer~\cite{Ugeda2016NbSe2}. Monolayer 1H-NbSe$_2$ also hosts Ising superconductivity with a large in-plane critical field~\cite{Xi2016Ising}, underscoring the interplay among lattice, spin-orbit, and pairing degrees of freedom.

The microscopic origin of the CDW in NbSe$_2$ has been debated for decades. Early proposals emphasized Fermi-surface nesting or saddle-point singularities, but subsequent theory and experiment have shown that nesting alone is insufficient~\cite{Johannes2008CDW,Rossnagel2011CDW}. Inelastic x-ray scattering reveals an extended phonon collapse near the CDW wave vector, supporting a momentum-dependent electron-phonon coupling (EPC) mechanism~\cite{Weber2011NbSe2}. Angle-resolved photoemission spectroscopy (ARPES) further establishes strong EPC signatures, CDW-related gaps, and Fermi-surface arcs that persist above the bulk transition~\cite{Valla2004NbSe2,Borisenko2009NbSe2}. From a theoretical perspective, density functional theory calculations successfully capture the CDW instability in bulk NbSe$_2$ and reveal that dimensionality influences the soft-mode wave vector and electronic ground state in few-layer NbSe$_2$~\cite{Calandra2009NbSe2}. Subsequent first-principles studies have mapped CDW superconductivity competition and EPC in monolayer NbSe$_2$~\cite{Lian2018NbSe2,Zheng2019NbSe2,Xu2018NbSe2}. Because the harmonic phonon spectrum of both bulk 2H-NbSe$_2$ and monolayer 1H-NbSe$_2$ are unstable above $T_{\mathrm{CDW}}$, anharmonic renormalization that incorporates temperatures effects is indispensable. In particular, stochastic self-consistent harmonic approximation (SSCHA)~\cite{Bianco2020NbSe2} calculations demonstrate that ionic fluctuations stabilize the lattice and largely control $T_{\mathrm{CDW}}$, with only a weak intrinsic thickness dependence once anharmonicity is fully included. More recently, machine learning interatomic potentials have enabled large-scale molecular dynamics and anharmonic lattice dynamics simulations of CDW reconstructions and soft phonons in monolayer and few-layer NbSe$_2$~\cite{rivano2026nbse2,benic2025nbse2}.

Despite these advances, the role of dynamical anharmonic phonon scattering in CDW soft-mode behavior and its consequences for electronic excitations remain poorly understood. Existing theoretical treatments have predominantly relied on statically renormalized phonon frequencies within the quasiparticle approximation, as exemplified by the use of SSCHA~\cite{SSCHA} in Ref.~\cite{Bianco2020NbSe2} and self-consistent phonon theory (SCPH)~\cite{Werthamer1970,SCPH2015} in Ref.~\cite{yang2026charge}. Such approaches neglect the full frequency dependence of phonon self-energies, hereafter referred to as dynamical anharmonic effects~\cite{monacelli2025analyzing}. Near a CDW instability, strong dynamical anharmonicity can substantially shift and broaden phonon spectra, producing non-Lorentzian lineshapes~\cite{Weber2011NbSe2} that challenge the quasiparticle picture. Although full frequency-dependent phonon spectral functions have recently been shown to be important for thermal transport in strongly anharmonic materials~\cite{dangic2021origin,xia2025lattice}, their implications for CDW soft-mode dynamics, transition temperatures, and electronic excitations remain largely unexplored. In particular, dynamically renormalized soft phonons may significantly modify electron-phonon interactions and leave distinct signatures in electronic spectra. This raises a fundamental question: How does dynamical anharmonicity reshape CDW soft modes beyond static renormalization, and how do these changes influence electronic excitations?

\begin{figure}[htp]
    \includegraphics[width = 1.0\linewidth]{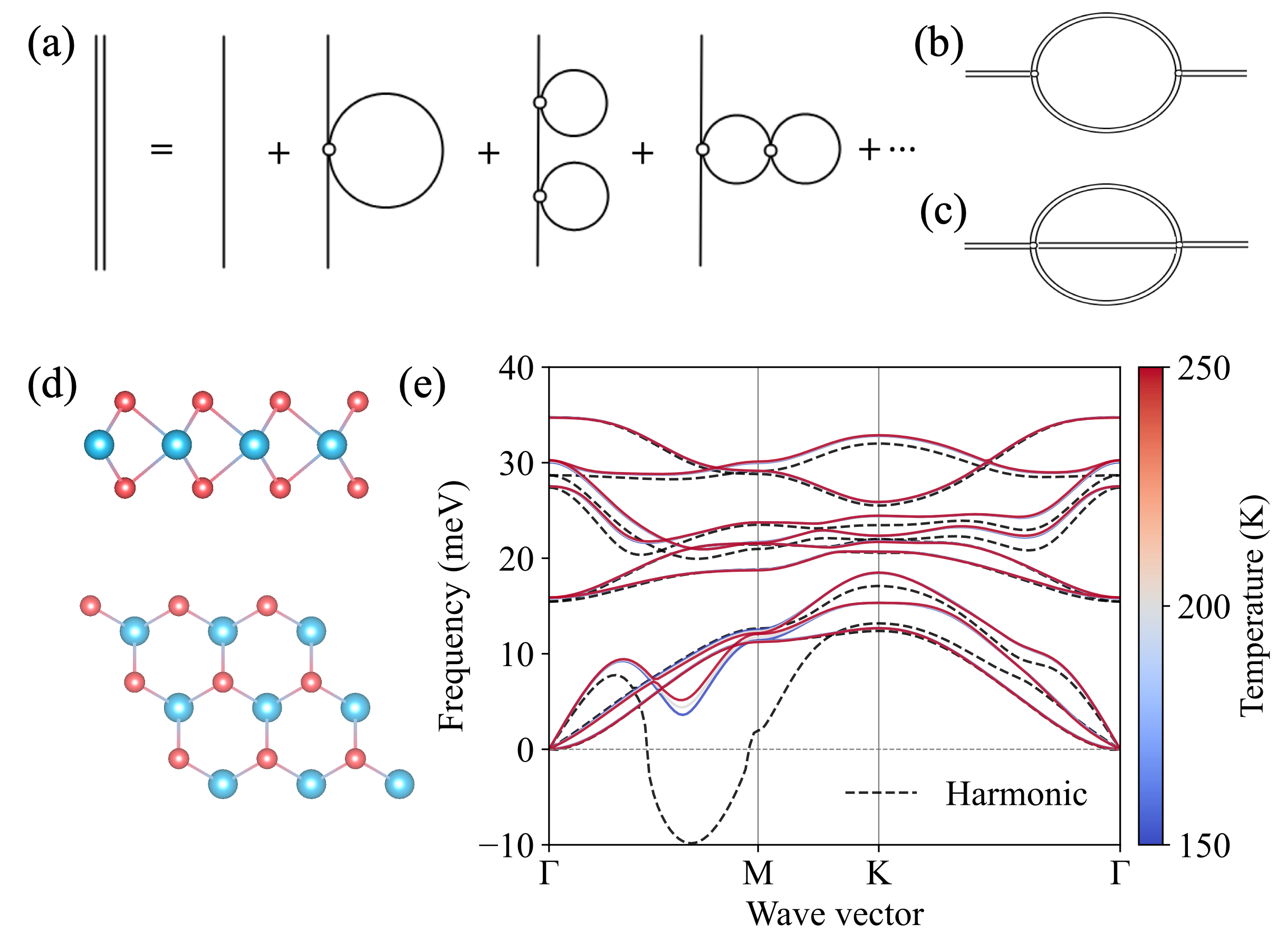}
    \caption{
    Crystal structure and anharmonic phonons in monolayer 1H-NbSe$_2$. (a) Diagrammatic representation of self-consistent phonon (SCPH) theory generated by quartic anharmonic loop insertions. Dynamical, frequency-dependent phonon self-energies from (b) the three-phonon (3ph) bubble and (c) the four-phonon (4ph) sunset diagrams built on top of SCPH. (d) Side and top views of the monolayer crystal structure with Nb in blue and Se in red. (e) Harmonic phonon dispersion (black dashed curves) compared with SCPH dispersions from 150 to 250~K.
    }
    \label{fig:scph}
\end{figure}

In this study, we address this question by investigating the interplay between dynamical anharmonicity, CDW soft modes, and electronic excitations in monolayer 1H-NbSe$_2$ near and above $T_{\rm CDW}$. First, we calculate frequency-dependent phonon spectral functions by combining SCPH theory with dynamical three-phonon (3ph) bubble and four-phonon (4ph) sunset self-energies. This approach enables us to elucidate how anharmonic interactions reshape the soft-mode spectra through frequency shifts and spectral broadening beyond the conventional quasiparticle approximation. Second, we examine how dynamical phonon renormalization influences the estimated CDW transition temperature, revealing the giant dynamical softening induced by 3ph and 4ph interactions. Third, we incorporate dynamically renormalized phonon frequencies and electron-phonon coupling matrix elements into Fan-Migdal electron self-energy calculations to investigate their impact on electronic spectral functions near the CDW transition. Our results demonstrate that dynamical anharmonicity substantially modifies both phonon and electronic excitations, establishing a crucial connection between beyond-quasiparticle lattice dynamics and electronic spectral signatures in two-dimensional CDW materials.

\begin{figure*}[htp]
    \includegraphics[width = 1.0\linewidth]{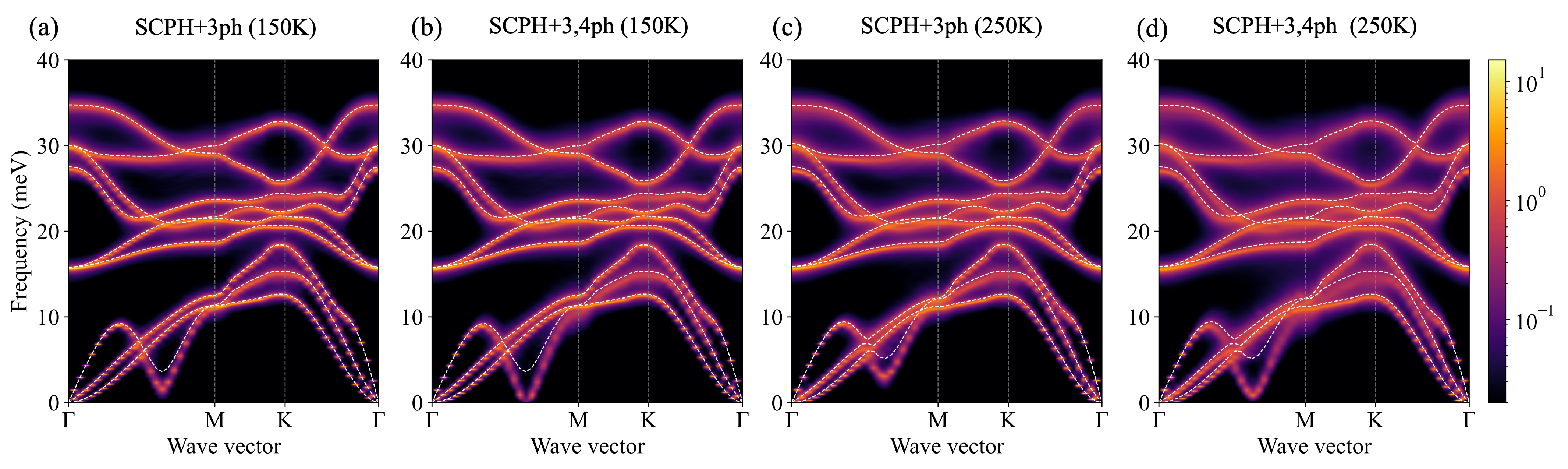}
    \caption{
    Phonon spectral functions of monolayer 1H-NbSe$_2$ along $\Gamma$--$M$--$K$--$\Gamma$. The color maps show the full frequency-dependent spectral functions obtained with SCPH frequencies plus the dynamical three-phonon bubble self-energy [(a) 150~K; (c) 250~K], and with both the three-phonon (3ph) bubble and four-phonon (4ph) sunset self-energies [(b) 150~K; (d) 250~K]. White dashed white curves mark the corresponding SCPH phonon frequencies.
    }
    \label{fig:phspec}
\end{figure*}

\section{Theory and Methods}

We begin by briefly outlining the diagrammatic Green's function approach to anharmonic lattice dynamics developed in our previous work~\cite{xia2025lattice}, as illustrated in Fig.~\ref{fig:scph}(a)–(c). The central feature of this approach is the consistent treatment of static and dynamical anharmonic effects on equal footing, while avoiding double counting. Specifically, static anharmonic renormalization establishes temperature-dependent reference phonon frequencies, which serve as the basis for evaluating dynamical phonon scattering. The latter introduces a frequency-dependent self-energy that governs phonon linewidths, spectral peak shifts, and non-Lorentzian lineshapes. Both contributions are essential: without static renormalization, the harmonic reference remains dynamically unstable, whereas neglecting dynamical scattering reduces the soft-mode spectral function to a $\delta$ function centered at the statically renormalized frequency.

The calculation proceeds in two stages. First, we obtain temperature-dependent phonon frequencies from SCPH theory~\cite{Werthamer1970,Errea2011,SCPH2015} by self-consistently dressing the phonon propagator with the quartic loop diagram [Fig.~\ref{fig:scph}(a)]. This Hartree-Fock-like renormalization~\cite{tadano2022first,xia2020particlelike,xia2025lattice,xiao2023anharmonic} incorporates the leading thermal and zero-point contributions of quartic anharmonicity and yields positive SCPH frequencies $\omega_{\lambda}^{\mathrm{SCPH}}(T)$ even when the harmonic spectrum is unstable. Second, we evaluate dynamical, frequency-dependent self-energies $\Sigma_{\lambda}(\omega)$ on these SCPH propagators from the 3ph bubble [Fig.~\ref{fig:scph}(b)] and 4ph sunset [Fig.~\ref{fig:scph}(c)] diagrams in the diagonal approximation,
\begin{equation}
\Sigma_{\lambda}(\omega)\approx\Sigma_{\lambda}^{3\mathrm{ph}}(\omega)+\Sigma_{\lambda}^{4\mathrm{ph}}(\omega),
\label{eq:sigma}
\end{equation}
where $\lambda$ labels the phonon wave vector and branch. The 3ph bubble diagram generates cubic anharmonic scattering channels, while the 4ph sunset diagram accounts for quartic processes that become increasingly important at elevated temperature and for soft modes with large vibrational amplitudes~\cite{xia2020microscopic,yang2026charge}. The interacting phonon spectral function $\mathcal{B}_{\lambda}(\omega)$ follows from the imaginary part of the dressed Green's function,
\begin{equation}
\mathcal{B}_{\lambda}(\omega)=
-\frac{1}{\pi}\,
\mathrm{Im}\,
\frac{2\omega_{\lambda}}
{\omega^2-\omega_{\lambda}^2-2\omega_{\lambda}\Sigma_{\lambda}(\omega)},
\label{eq:Aph}
\end{equation}
with $\omega_{\lambda}$ taken as the SCPH frequency. The real and imaginary parts of $\Sigma_{\lambda}(\omega)$ simultaneously broaden and shift the spectral peaks. Explicit expressions for $\Sigma_{\lambda}^{3\mathrm{ph}}$ and $\Sigma_{\lambda}^{4\mathrm{ph}}$ can be found in Ref.~\cite{xia2025lattice}. Computational details regarding the force constant extraction, Brillouin-zone sampling, and convergence tests are provided in Supplementary Materials~\cite{SM}.

Having established the phonon framework, we compute electronic spectra within the Electron-Phonon Wannier (EPW) framework~\cite{Giustino2007,Noffsinger20102140,Ponce2016,giustino2017electron,lee2023electron}. Because harmonic phonons of monolayer 1H-NbSe$_2$ exhibit imaginary soft modes, a direct harmonic EPW calculation is ill-defined for temperature-dependent spectra. We therefore supply anharmonically renormalized interatomic force constants to EPW~\cite{yin2025impact}, from which renormalized phonon frequencies and eigenvectors are obtained. The corresponding electron-phonon matrix elements are then Wannier-interpolated onto dense Brillouin-zone grids. Two levels of phonon renormalization are considered: (i) static force constants from SCPH, and (ii) dynamical quasiparticle-renormalized force constants constructed from the peaks of the full phonon spectral functions including 3ph and 4ph interactions (SCPH+3ph and SCPH+3,4ph). The temperature-dependent electron self-energy is evaluated from the Fan-Migdal electron self-energy diagram using these renormalized phonons and interpolated matrix elements, and the electronic spectral function is obtained from the resulting complex self-energy~\cite{giustino2017electron,ponce2020first}. Computational details regarding electronic structure simulation, electron-phonon Wannier interpolation, and self-energy calculations are provided in Supplementary Materials~\cite{SM}.


\section{Results and Discussion}

We begin with the analysis of lattice dynamical stability of monolayer 1H-NbSe$_2$. The monolayer consists of a Nb plane sandwiched by Se atoms in trigonal prismatic coordination [Fig.~\ref{fig:scph}(d)]. In the harmonic approximation, an acoustic branch exhibits imaginary frequencies along $\Gamma$-$M$, with the most unstable modes concentrated near $\tfrac{2}{3}\,\Gamma$-$M$ and reaching approximately $-10$~meV [Fig.~\ref{fig:scph}(e)]. The location of this wavevector coincides with the established CDW ordering vector of the $3\times3$ reconstruction in monolayer 1H-NbSe$_2$, and is consistent with earlier first-principles phonon calculations of CDW-related soft phonons~\cite{Calandra2009NbSe2,Bianco2020NbSe2}. By including static anharmonic renormalization effects, SCPH removes the imaginary branch and restores a stable dispersion from 150 to 250~K. The soft mode near $\tfrac{2}{3}\,\Gamma$-$M$ hardens systematically with increasing temperature, while other acoustic modes and higher-energy optical branches exhibit comparatively weaker thermal shifts. These trends agree with the picture that ionic fluctuations, rather than purely electronic thermal smearing, dominate the melting of CDW order in NbSe$_2$~\cite{Bianco2020NbSe2,yang2026charge}. Our results show that, even at 150~K, just above the reported monolayer $T_{\mathrm{CDW}}$~\cite{Xi2015monolayerCDW}, the SCPH soft-mode frequency remains finite, providing a well-defined starting point for dynamical self-energy calculations.

\begin{figure*}[!t]
    \includegraphics[width = 1.0\linewidth]{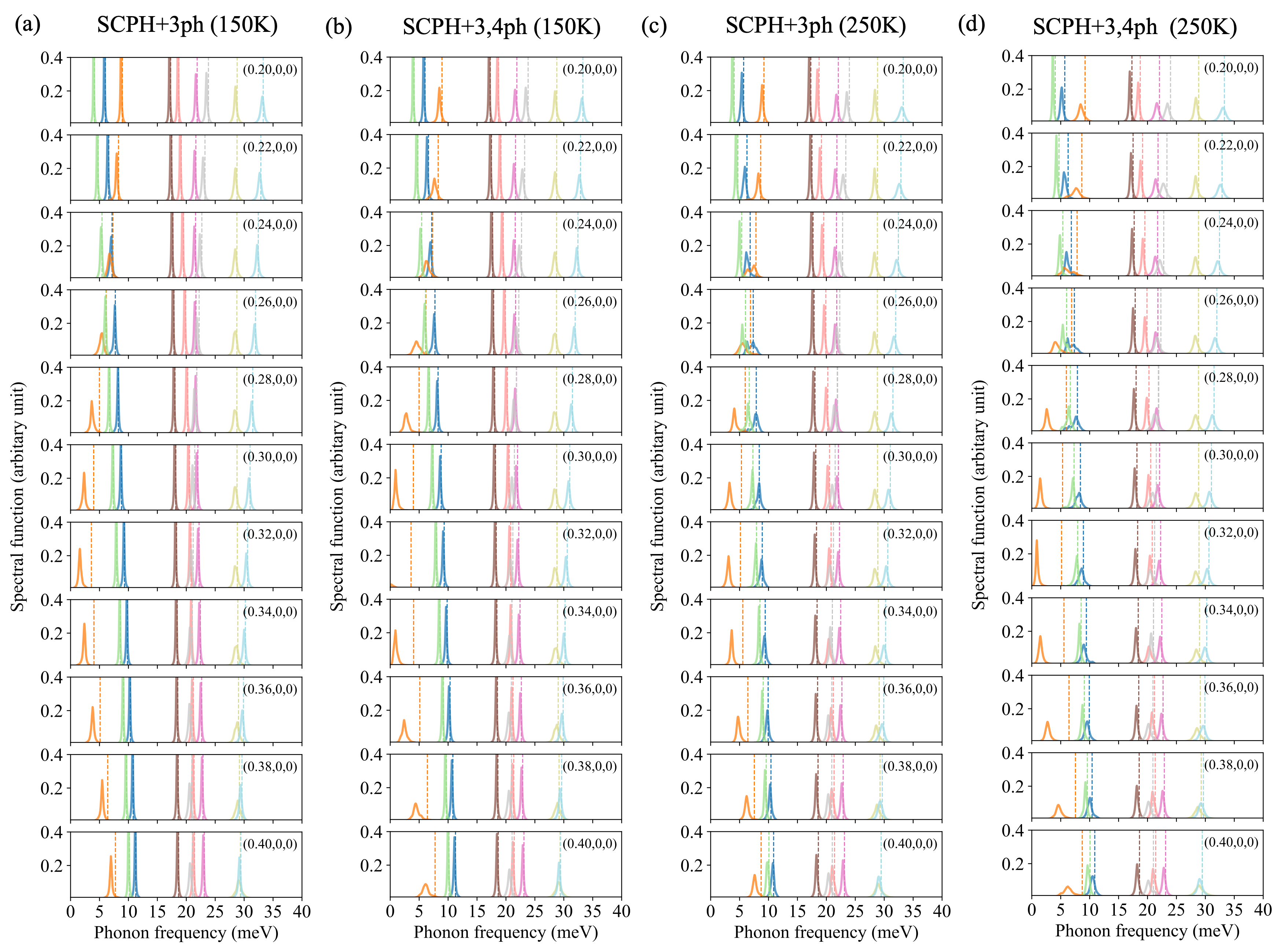}
    \caption{
    Mode-resolved phonon spectral functions of monolayer 1H-NbSe$_2$ along $\Gamma$--$M$, corresponding to the same calculations as in Fig.~\ref{fig:phspec}. Columns show results at (a,b) 150~K and (c,d) 250~K with the dynamical three-phonon (3ph) bubble self-energy [(a),(c)] and with both the 3ph bubble and four-phonon (4ph) sunset self-energies [(b),(d)]. Rows correspond to phonon wave vectors $(\xi,0,0)$ with $\xi=0.20$-$0.40$. Solid curves are the frequency-dependent spectral functions of individual phonon modes; vertical dashed lines mark the corresponding SCPH frequencies. Modes are colored according to their vibrational character from eigenvectors, as assigned by an analysis of phonon band connectivity using Phonopy software~\cite{phonopy-phono3py-JPCM, phonopy-phono3py-JPSJ}.
    }
    \label{fig:phmode}
\end{figure*}

Going beyond static SCPH frequencies, Fig.~\ref{fig:phspec} presents the phonon spectral functions obtained by dressing the SCPH propagator with dynamical 3ph bubble and 4ph sunset self-energies. 
We compare SCPH+3ph and SCPH+3,4ph spectra at 150 and 250~K to separate the effects from higher-order 4ph interactions.
At 150~K with only 3ph interactions [Fig.~\ref{fig:phspec}(a)], strong dynamical softening appears at the CDW phonon wave vectors near $\tfrac{2}{3}\,\Gamma$-$M$, with the soft-mode spectral peak lying well below the SCPH reference, whereas the remaining branches exhibit only slight dynamical softening and remain comparatively sharp. Including additional 4ph scattering [Fig.~\ref{fig:phspec}(b)] further enhances the dynamical softening of CDW-related phonon modes, driving the mode near $\tfrac{2}{3}\,\Gamma$–$M$ below zero meV and signaling a dynamical lattice instability. Meanwhile, 4ph interactions induce an overall broadening of the phonon spectrum. Notably, the lowest-energy phonon modes near the $M$ points exhibit highly diffuse spectral features [Fig.~\ref{fig:phspec}(b)], suggesting a potential breakdown of the phonon quasiparticle picture.

Similar dynamical softening persists at an elevated temperature of 250~K [Figs.~\ref{fig:phspec}(c) and (d)]. A comparison of these panels reveals that 4ph interactions play an even more prominent role in dynamically softening the CDW-related modes at higher temperatures, consistent with the stronger temperature dependence of higher-order phonon–phonon scattering processes. Furthermore, the elevated temperature leads to substantial broadening of the phonon spectral functions, with diffuse spectral features extending across much of the Brillouin zone [Figs.~\ref{fig:phspec} (d)], suggesting a widespread breakdown of the phonon quasiparticle picture. This dynamical softening is qualitatively distinct from the static SCPH hardening with temperature: while SCPH frequencies themselves increase upon heating, the peaks of spectral functions can move downward relative to those SCPH frequencies because the real part of the frequency-dependent phonon self-energy, $\mathrm{Re}\,\Sigma_{\lambda}(\omega)$, is frequency dependent. Experimentally, the extended phonon collapse observed by inelastic x-ray scattering in bulk NbSe$_2$ already hinted that soft mode spectral weight is strongly renormalized over a broad momentum and energy window~\cite{Weber2011NbSe2}. Fig.~\ref{fig:phspec} shows that an analogous beyond-quasiparticle renormalization emerges in the monolayer 1H-NbSe$_2$ once dynamical 3ph bubble and 4ph sunset diagrams are retained.

The mode-resolved spectra along $\Gamma$--$M$ in Fig.~\ref{fig:phmode} make this physics quantitative. We focus on phonon wave vectors $(\xi,0,0)$ with $\xi=0.20-0.40$ because this window encompasses the CDW ordering vector near $\tfrac{2}{3}\,\Gamma$-$M$. Phonon branches are colored according to eigenvector continuity from a careful analysis of band connectivity, enabling unambiguous tracking of the soft mode across neighboring phonon wave vectors. Within our $50\times50\times1$ phonon mesh, the lowest CDW-relevant mode is identified at $(0.32,0,0)$, the grid point closest to $\tfrac{2}{3}\,\Gamma$--$M$. At 150~K with only 3ph interactions [Fig.~\ref{fig:phmode}(a)], this soft mode remains a sharp, well-defined peak well below the corresponding SCPH frequency (dashed vertical lines), while higher-lying branches stay comparatively narrow and close to their SCPH references. Adding 4ph scattering [Fig.~\ref{fig:phmode}(b)] not only broadens but also significantly softens the low-lying CDW soft modes. Note that the such strong softening even eliminates the lowest-lying soft mode at $(0.32,0,0)$ [Fig.~\ref{fig:phmode}(b)], rendering the lattice unstable again. Importantly, the pronounced dynamical softening induced by 3ph and 4ph interactions remains largely confined to the CDW-related modes, highlighting the mode-selective nature of these anharmonic effects. 

At 250~K, the increase in the SCPH reference frequencies leads to a hardening of the lowest-energy phonon spectral peaks near the CDW wave vectors in both SCPH+3ph [Fig.~\ref{fig:phmode}(c)] and SCPH+3,4ph [Fig.~\ref{fig:phmode}(d)] calculations. Interestingly, the lowest-energy soft mode at $(0.32,0,0)$ reemerges at positive frequency in the SCPH+3,4ph calculation, signaling the restoration of dynamical lattice stability. Compared with the mode-resolved spectral functions at 150~K, those at 250~K exhibit substantially enhanced broadening, with pronounced asymmetric and non-Lorentzian lineshapes emerging across the Brillouin zone. These non-Lorentzian spectral features highlight the limitations of describing phonon dynamics solely in terms of statically renormalized frequencies.

\begin{figure}[htp]
    \includegraphics[width = 1.0\linewidth]{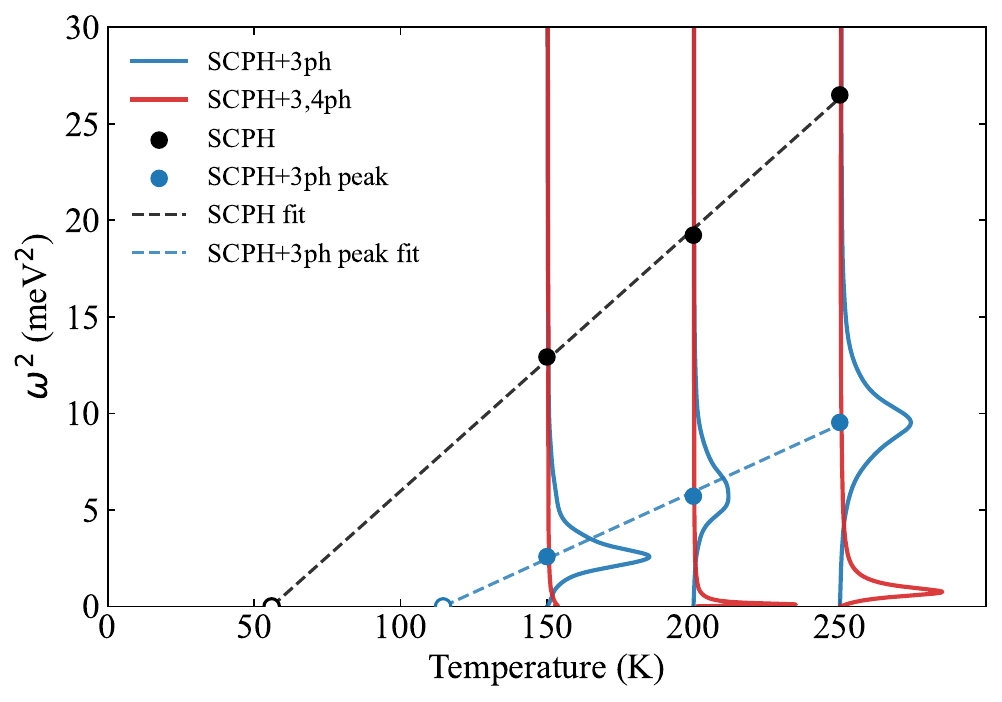}
    \caption{
    Temperature dependence of the soft-mode frequency squared, $\omega^2$, used to estimate CDW transition temperature of monolayer 1H-NbSe$_2$. Black and blue circles show $\omega^2$ from SCPH and from the peak of the full frequency-dependent phonon spectral function with three-phonon bubble interactions (SCPH+3ph), respectively. Dashed lines are linear fits extrapolated to $\omega^2=0$. Overlayed spectral functions at selected temperatures compare SCPH+3ph (blue) and SCPH+3,4ph (red) results.
    }
    \label{fig:trans}
\end{figure}

To connect these spectral features to the CDW transition, we analyze the temperature dependence of the soft-mode frequency [Fig.~\ref{fig:trans}]. The estimation of soft-mode $T_{\mathrm{CDW}}$ is obtained by identifying the maximum of the spectral function of the lowest-lying mode at (0.32,0,0) and extrapolating $\omega_{\mathrm{peak}}^2(T)$ linearly to zero. Both SCPH frequencies and SCPH+3ph spectral-peak frequencies soften approximately linearly upon cooling, as expected for a second-order displacive instability~\cite{Weber2011NbSe2,Bianco2020NbSe2}. Linear extrapolation of $\omega^2$ to zero yields $T_{\mathrm{CDW}}\approx65$~K from SCPH alone. Using instead the dynamically softened SCPH+3ph peak increases the estimate to $\approx115$~K, a substantial upward revision of about 77\% arising solely from frequency-dependent 3ph interactions. Including 4ph interactions further enhances the CDW transition temperature. Specifically, the spectral peak associated with the CDW soft mode shifts entirely below zero meV at 150~K, begins to reemerge at positive energies at 200~K, and fully recovers at 250~K, suggesting a CDW transition temperature between 150 and 200~K.

We emphasize that our objective here is not to precisely predict the absolute CDW transition temperature, but rather to elucidate the influence of dynamical effects on soft-mode behavior. Unlike the free-energy-based determination of $T_{\mathrm{CDW}}$ within SSCHA~\cite{Bianco2020NbSe2}, our estimate is obtained by extrapolating the squared soft-mode frequency to zero, which assumes a purely displacive, continuous phase transition. In reality, the CDW transition may involve intertwined displacive and order-disorder mechanisms~\cite{kumbhakar2026dynamics}, potentially limiting the applicability of our approach. Furthermore, the estimated $T_{\mathrm{CDW}}$ may be sensitive to the exchange-correlation functional~\cite{yang2026charge}, thermal expansion, and dynamical electron-phonon interactions~\cite{exists2026giant}, which are not fully accounted for in the present treatment and may contribute to deviations from experimental observations~\cite{Xi2015monolayerCDW,Ugeda2016NbSe2}. Nevertheless, Fig.~\ref{fig:trans} highlights the significant role of dynamical anharmonicity: relative to static SCPH frequencies, frequency-dependent 3ph and 4ph interactions substantially soften the CDW mode and increase the inferred instability temperature. 

\begin{figure*}[htp]
    \includegraphics[width = 1.0\linewidth]{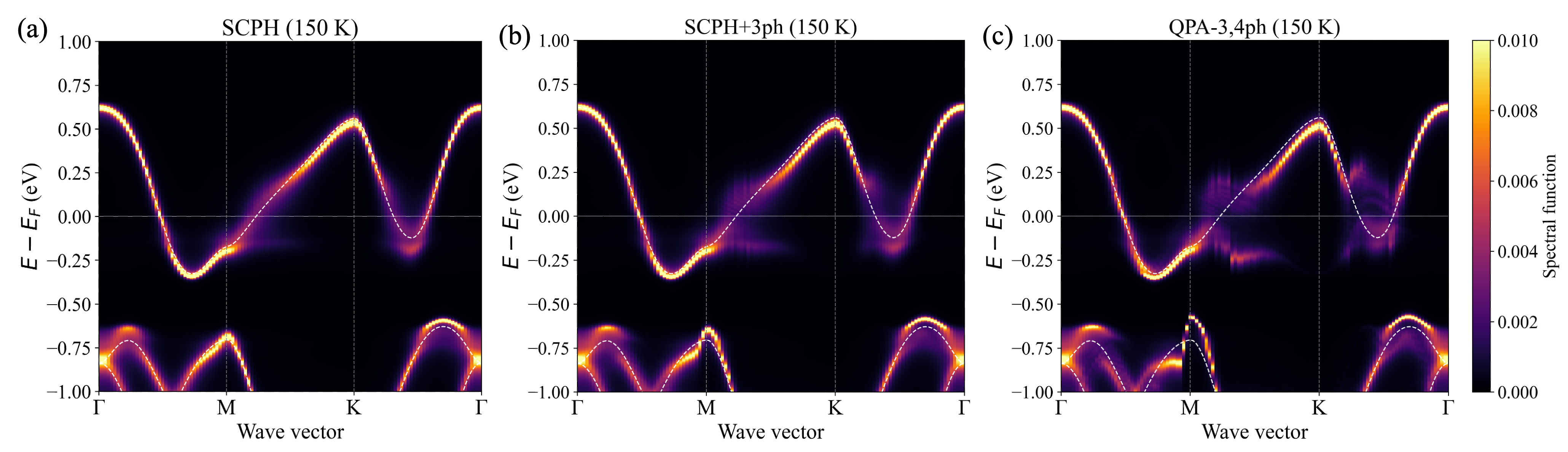}
    \caption{
    Electronic spectral function of monolayer 1H-NbSe$_2$ near Fermi level along $\Gamma$--$M$--$K$--$\Gamma$, calculated using (a) SCPH frequencies at 150~K, (b) the quasiparticle approximation based on SCPH+3ph at 150~K, and (c) SCPH+3,4ph at 150~K, in which phonon frequencies are taken from the identified quasiparticle peaks of the corresponding phonon spectral functions.
    Energies are referenced to the Fermi level, $E_F$.
    }
    \label{fig:elspec}
\end{figure*}

Finally, we investigate how anharmonic phonon interactions influence the electronic spectral function at 150~K, near the reported CDW transition temperature of monolayer NbSe$_2$ [Fig.~\ref{fig:elspec}]. For each case, the corresponding renormalized force constants (see Fig.S1 for a comparison of phonon dispersion obtained from SCPH, SCPH+3ph, and SCPH+3,4ph in Supplementary Materials~\cite{SM}) are used to interpolate the electron-phonon coupling matrix elements within the EPW framework, which, together with the renormalized phonon frequencies, enter the Fan–Migdal self-energy calculations. As shown in Fig.~\ref{fig:elspec}(a), the SCPH-based electronic spectral function exhibits well-defined quasiparticle bands, in good agreement with previous first-principles calculations~\cite{Calandra2009NbSe2,Ugeda2016NbSe2}. Notably, kink-like features emerge near the $M$ point, accompanied by satellite structures along the $K$-$\Gamma$ direction, revealing appreciable electron-phonon renormalization. Including 3ph interactions [Fig.~\ref{fig:elspec}(b)] further enhances these features, producing more pronounced kinks, stronger satellite spectral weight, and additional quasiparticle broadening. This enhancement can be attributed to the 3ph-induced reduction in the CDW-related soft-mode frequencies, which increases their thermal populations and consequently strengthens electron-phonon scattering. The pronounced momentum selectivity of these spectral features further highlights the strongly momentum-dependent nature of the electron-phonon coupling matrix elements (see Fig.S3 in Supplementary Materials~\cite{SM}) in NbSe$_2$~\cite{Valla2004NbSe2,flicker2015charge,Zheng2019NbSe2,das2023electron}.

A qualitatively different behavior emerges upon incorporating additional 4ph interactions through the SCPH+3,4ph renormalized force constants [Fig.~\ref{fig:elspec}(c)]. Pronounced quasiparticle decoherence develops across substantial portions of the Brillouin zone, particularly along $M$–$K$ and $K$–$\Gamma$, where well-defined quasiparticle bands become strongly broadened and spectral weight is redistributed into diffuse, incoherent features. This behavior is closely associated with the exceptionally low frequencies of the CDW-related soft modes, which approach zero under 4ph-induced dynamical softening. The resulting diverging thermal phonon populations substantially enhance low-energy electron–phonon scattering and the Fan–Migdal self-energy, leading to pronounced quasiparticle broadening and loss of coherence. These results demonstrate that the progressive softening of CDW-related phonons, from SCPH to SCPH+3ph and SCPH+3,4ph, systematically amplifies electronic many-body effects, evolving from enhanced kinks and satellite features to pronounced quasiparticle decoherence. This underscores the critical interplay between higher-order phonon anharmonicity and momentum-dependent electron-phonon coupling in shaping the electronic spectra near a CDW instability~\cite{fragkos2026phonon}, even without explicitly accounting for CDW-induced electronic band reconstruction.

Before concluding, we highlight several limitations of the present approach and promising directions for future research. First, our evaluation of the dynamical phonon self-energy remains perturbative. Finite-temperature renormalization of anharmonic interaction vertices and a self-consistent treatment of phonon spectral functions may become essential in strongly anharmonic systems~\cite{xia2026self}. Second, the present calculations do not account for feedback between the electronic and phononic subsystems. Such feedback may be important because electron-phonon interactions modify phonon spectral functions, which in turn influence electronic self-energies and spectral properties~\cite{lihm2024self,exists2026giant,erhardt2026higher}. Third, in systems exhibiting strong dynamical anharmonicity and electron-phonon coupling, an accurate description may require incorporating the full frequency dependence of both phonon and electron spectral functions into electron-phonon self-energy calculations, beyond conventional quasiparticle-based approximations~\cite{lihm2026beyond,xia2026self}. Addressing these challenges represents promising directions toward a unified, self-consistent description of coupled electron-phonon dynamics in CDW materials and other strongly anharmonic systems.

\section{Conclusion}

In summary, we demonstrate that dynamical phonon anharmonicity plays a crucial role in shaping CDW instabilities and their electronic signatures in monolayer 1H-NbSe$_2$. By incorporating frequency-dependent 3ph and 4ph self-energies beyond static self-consistent phonon theory, we reveal pronounced softening of CDW-related modes, accompanied by substantial spectral broadening and non-Lorentzian lineshapes that challenge the phonon quasiparticle picture. Notably, 4ph interactions further enhance the CDW instability, yielding an estimated transition temperature between 150 and 200~K. Incorporating the dynamically renormalized phonons and electron-phonon coupling into Fan-Migdal calculations reveals enhanced electronic band kinks and satellite features due to 3ph-induced softening, while additional 4ph interactions drive pronounced momentum-dependent quasiparticle decoherence as the soft-mode frequencies approach zero. These findings highlight the critical interplay between higher-order phonon anharmonicity and momentum-dependent electron-phonon coupling in governing CDW instabilities and electronic quasiparticle dynamics in two-dimensional quantum materials.

\begin{acknowledgments}
\textbf{Acknowledgments:} Y. X. acknowledges support from the U.S. National Science Foundation through awards CBET-2445361, DMR-2532261, and DMR-2317008, as well as computing resources provided by Bridges2 at Pittsburgh Supercomputing Center through ACCESS allocations mat220006p and mat220008p. ACCESS is supported by National Science Foundation grants 2138259, 2138286, 2138307, 2137603, and 2138296. Y. X. is grateful to Z. J. W. for encouragement and support during the preparation of this manuscript.
\end{acknowledgments}


\bibliography{CuSbS}

@article{kumbhakar2026dynamics,
  title={Dynamics of Charge-Density-Wave puddles in 2H-NbSe2},
  author = {Kumbhakar, Shreya and Esposito, Marina and Kumar N M, Anjan and Feng, Liwen and Confalone, Tommaso and Gerega, Vasilisa and Lo Sardo, Flavia and Alam, Rafiqul and Masarotti, Davide and Tafuri, Francesco and Bohm, Thomas and Abdel-Hafiez, Mahmoud and Chandra, Sushmita and Felser, Claudia and Nielsch, Kornelius and Poccia, Nicola and Kaiser, Stefan and Haider, Golam},
  journal = {Nano Letters},
    volume = {26},
    number = {41},
    pages = {14376-14383},
    year = {2026},
    month = {10},
    issn = {1530-6984},
    doi = {10.1021/acs.nanolett.6c03967},
    url = {https://doi.org/10.1021/acs.nanolett.6c03967}
}

@article{phonopy-phono3py-JPCM,
  author  = {Togo, Atsushi and Chaput, Laurent and Tadano, Terumasa and Tanaka, Isao},
  title   = {Implementation strategies in phonopy and phono3py},
  journal = {J. Phys. Condens. Matter},
  volume  = {35},
  number  = {35},
  pages   = {353001},
  year    = {2023},
  doi     = {10.1088/1361-648X/acd831}
}

@article{phonopy-phono3py-JPSJ,
  author  = {Togo, Atsushi},
  title   = {First-principles Phonon Calculations with Phonopy and Phono3py},
  journal = {J. Phys. Soc. Jpn.},
  volume  = {92},
  number  = {1},
  pages   = {012001},
  year    = {2023},
  doi     = {10.7566/JPSJ.92.012001}
  }

@article{alshemi2025two,
  title={Two characteristic contributions to the superconducting state of 2H-NbSe2},
  author={Alshemi, A and Forgan, EM and Hiess, A and Cubitt, R and White, JS and Schmalzl, K and Blackburn, E},
  journal={Physical Review Letters},
  volume={134},
  number={11},
  pages={116001},
  year={2025},
  doi = {10.1103/PhysRevLett.134.116001},
  publisher={APS}
}

@article{xia2020particlelike,
  title = {Particlelike Phonon Propagation Dominates Ultralow Lattice Thermal Conductivity in Crystalline ${\mathrm{Tl}}_{3}{\mathrm{VSe}}_{4}$},
  author = {Xia, Yi and Pal, Koushik and He, Jiangang and Ozoli\ifmmode \mbox{\c{n}}\else \c{n}\fi{}\ifmmode \check{s}\else \v{s}\fi{}, Vidvuds and Wolverton, Chris},
  journal = {Phys. Rev. Lett.},
  volume = {124},
  issue = {6},
  pages = {065901},
  numpages = {7},
  year = {2020},
  month = {Feb},
  publisher = {American Physical Society},
  doi = {10.1103/PhysRevLett.124.065901},
  url = {https://link.aps.org/doi/10.1103/PhysRevLett.124.065901}
}

@article{ponce2020first,
  title={First-principles calculations of charge carrier mobility and conductivity in bulk semiconductors and two-dimensional materials},
  author={Ponc{\'e}, Samuel and Li, Wenbin and Reichardt, Sven and Giustino, Feliciano},
  journal={Reports on Progress in Physics},
  volume={83},
  number={3},
  pages={036501},
  year={2020},
  publisher={IOP Publishing},
  url={https://iopscience.iop.org/article/10.1088/1361-6633/ab6a43}
}

@article{erhardt2026higher,
  title={Higher-order nonadiabaticity governs the temperature dependence of the phonon spectrum},
  author={Erhardt, Nina Girotto and Ponc{\'e}, Samuel},
  journal={arXiv preprint arXiv:2608.14882},
  year={2026}
}

@article{lee2023electron,
  title={Electron--phonon physics from first principles using the EPW code},
  author={Lee, Hyungjun and Ponc{\'e}, Samuel and Bushick, Kyle and Hajinazar, Samad and Lafuente-Bartolome, Jon and Leveillee, Joshua and Lian, Chao and Lihm, Jae-Mo and Macheda, Francesco and Mori, Hitoshi and others},
  journal={npj Computational Materials},
  volume={9},
  number={1},
  pages={156},
  year={2023},
  publisher={Nature Publishing Group UK London},
  doi={10.1038/s41524-023-01107-3}
}

@article{xiao2023anharmonic,
  title={Anharmonic phonon behavior via irreducible derivatives: Self-consistent perturbation theory and molecular dynamics},
  author={Xiao, Enda and Marianetti, Chris A},
  journal={Physical Review B},
  volume={107},
  number={9},
  pages={094303},
  year={2023},
  publisher={APS}
}

@article{dangic2021origin,
  title={The origin of the lattice thermal conductivity enhancement at the ferroelectric phase transition in GeTe},
  author={Dangic, Dorde and Hellman, Olle and Fahy, Stephen and Savic, Ivana},
  journal={npj Computational Materials},
  volume={7},
  number={1},
  pages={57},
  year={2021},
  publisher={Nature Publishing Group UK London},
  url={https://www.nature.com/articles/s41524-021-00523-7}
}

@article{SCPH2015,
  title = {Self-consistent phonon calculations of lattice dynamical properties in cubic ${\mathrm{SrTiO}}_{3}$ with first-principles anharmonic force constants},
  author = {Tadano, Terumasa and Tsuneyuki, Shinji},
  journal = {Phys. Rev. B},
  volume = {92},
  issue = {5},
  pages = {054301},
  numpages = {10},
  year = {2015},
  month = {Aug},
  publisher = {American Physical Society},
  doi = {10.1103/PhysRevB.92.054301},
  url = {https://link.aps.org/doi/10.1103/PhysRevB.92.054301}
}

@article{lihm2026beyond,
  title = {Beyond-Quasiparticle Transport with Vertex Correction: Self-Consistent Ladder Formalism for Electron-Phonon Interactions},
  author = {Lihm, Jae-Mo and Ponc\'e, Samuel},
  journal = {Phys. Rev. X},
  volume = {16},
  issue = {1},
  pages = {011008},
  numpages = {47},
  year = {2026},
  month = {Jan},
  publisher = {American Physical Society},
  doi = {10.1103/mtjf-ylf7},
  url = {https://link.aps.org/doi/10.1103/mtjf-ylf7}
}

@article{lihm2024self,
  title = {Self-consistent electron lifetimes for electron-phonon scattering},
  author = {Lihm, Jae-Mo and Ponc\'e, Samuel and Park, Cheol-Hwan},
  journal = {Phys. Rev. B},
  volume = {110},
  issue = {12},
  pages = {L121106},
  numpages = {7},
  year = {2024},
  month = {Sep},
  publisher = {American Physical Society},
  doi = {10.1103/PhysRevB.110.L121106},
  url = {https://link.aps.org/doi/10.1103/PhysRevB.110.L121106}
}

@article{xia2025lattice,
  title = {Lattice thermal transport beyond the quasiparticle approximation: Nontrivial spectral competition between three- and four-phonon interactions},
  author = {Xia, Yi},
  journal = {Phys. Rev. B},
  volume = {112},
  issue = {24},
  pages = {L241201},
  numpages = {7},
  year = {2025},
  month = {Dec},
  publisher = {American Physical Society},
  doi = {10.1103/r3sd-ftgb},
  url = {https://link.aps.org/doi/10.1103/r3sd-ftgb}
}

@article{giustino2017electron,
  title = {Electron-phonon interactions from first principles},
  author = {Giustino, Feliciano},
  journal = {Rev. Mod. Phys.},
  volume = {89},
  issue = {1},
  pages = {015003},
  numpages = {63},
  year = {2017},
  month = {Feb},
  publisher = {American Physical Society},
  doi = {10.1103/RevModPhys.89.015003},
  url = {https://link.aps.org/doi/10.1103/RevModPhys.89.015003}
}

@article{flicker2015charge,
  title={Charge order from orbital-dependent coupling evidenced by NbSe2},
  author={Flicker, Felix and Van Wezel, Jasper},
  journal={Nature communications},
  volume={6},
  number={1},
  pages={7034},
  year={2015},
  publisher={Nature Publishing Group UK London},
  doi={10.1038/ncomms8034}
}

@article{das2023electron,
  title={Electron-phonon coupling and spin fluctuations in the Ising superconductor NbSe2},
  author={Das, Suvadip and Paudyal, Hari and Margine, Elena R and Agterberg, DF and Mazin, II},
  journal={npj Computational Materials},
  volume={9},
  number={1},
  pages={66},
  year={2023},
  publisher={Nature Publishing Group UK London},
  doi={10.1038/s41524-023-01017-4}
}

@article{tadano2022first,
  title = {First-Principles Phonon Quasiparticle Theory Applied to a Strongly Anharmonic Halide Perovskite},
  author = {Tadano, Terumasa and Saidi, Wissam A.},
  journal = {Phys. Rev. Lett.},
  volume = {129},
  issue = {18},
  pages = {185901},
  numpages = {7},
  year = {2022},
  month = {Oct},
  publisher = {American Physical Society},
  doi = {10.1103/PhysRevLett.129.185901},
  url = {https://link.aps.org/doi/10.1103/PhysRevLett.129.185901}
}

@article{fragkos2026phonon,
  title={Phonon-induced pseudogap phase in TiSe $ \_2$},
  author={Fragkos, Sotirios and Erhardt, Nina Girotto and Symeonidou, Evgenia and Orio, Hibiki and Descamps, Dominique and Petit, St{\~A}{\v{S}}phane and Tsipas, Polychronis and Rossnagel, Kai and Schusser, Jakub and Dimoulas, Athanasios and others},
  journal={arXiv preprint arXiv:2606.31298},
  year={2026}
}

@article{monacelli2025analyzing,
  title={Analyzing the anharmonic phonon spectrum: Self-consistent approximation and temperature-dependent effective potential methods},
  author={Monacelli, Lorenzo},
  journal={Physical Review B},
  volume={112},
  number={1},
  pages={014109},
  year={2025},
  publisher={APS}
}

@misc{SM,
	Title = {Supplemental Materials contain details on the theory, computational methods and parameters, as well as convergence tests.}}

@article{Errea2011,
	Author = {Errea, Ion and Rousseau, Bruno and Bergara, Aitor},
	Doi = {10.1103/PhysRevLett.106.165501},
	Issue = {16},
	Journal = {Phys. Rev. Lett.},
	Month = {Apr},
	Numpages = {4},
	Pages = {165501},
	Publisher = {American Physical Society},
	Title = {Anharmonic Stabilization of the High-Pressure Simple Cubic Phase of Calcium},
	Url = {https://link.aps.org/doi/10.1103/PhysRevLett.106.165501},
	Volume = {106},
	Year = {2011}}

@article{xia2020microscopic,
  title = {Microscopic Mechanisms of Glasslike Lattice Thermal Transport in Cubic ${\mathrm{Cu}}_{12}{\mathrm{Sb}}_{4}{\mathrm{S}}_{13}$ Tetrahedrites},
  author = {Xia, Yi and Ozoli\ifmmode \mbox{\c{n}}\else \c{n}\fi{}\ifmmode \check{s}\else \v{s}\fi{}, Vidvuds and Wolverton, Chris},
  journal = {Phys. Rev. Lett.},
  volume = {125},
  issue = {8},
  pages = {085901},
  numpages = {7},
  year = {2020},
  month = {Aug},
  publisher = {American Physical Society},
  doi = {10.1103/PhysRevLett.125.085901},
  url = {https://link.aps.org/doi/10.1103/PhysRevLett.125.085901}
}

@article{SSCHA,
doi = {10.1088/1361-648X/ac066b},
url = {https://dx.doi.org/10.1088/1361-648X/ac066b},
year = {2021},
month = {jul},
publisher = {IOP Publishing},
volume = {33},
number = {36},
pages = {363001},
author = {Lorenzo Monacelli and Raffaello Bianco and Marco Cherubini and Matteo Calandra and Ion Errea and Francesco Mauri},
title = {The stochastic self-consistent harmonic approximation: calculating vibrational properties of materials with full quantum and anharmonic effects},
journal = {Journal of Physics: Condensed Matter},
}

@article{yin2025impact,
  title={Impact of anharmonicity on the carrier mobility of the Pb-free CsSnBr 3 perovskite},
  author={Yin, Junwen and Hellman, Olle and Ponc{\'e}, Samuel},
  journal={Physical Review B},
  volume={112},
  number={14},
  pages={L140303},
  year={2025},
  publisher={APS}
}

@article{Werthamer1970,
	Author = {Werthamer, N. R.},
	Doi = {10.1103/PhysRevB.1.572},
	Issue = {2},
	Journal = {Phys. Rev. B},
	Month = {Jan},
	Numpages = {0},
	Pages = {572--581},
	Publisher = {American Physical Society},
	Title = {Self-Consistent Phonon Formulation of Anharmonic Lattice Dynamics},
	Url = {https://link.aps.org/doi/10.1103/PhysRevB.1.572},
	Volume = {1},
	Year = {1970}}

@article{Ponce2016,
	Author = {S. Ponc{\'e} and E.R. Margine and C. Verdi and F. Giustino},
	Doi = {http://dx.doi.org/10.1016/j.cpc.2016.07.028},
	Issn = {0010-4655},
	Journal = {Computer Physics Communications},
	Pages = {-},
	Title = {EPW: Electron--phonon coupling, transport and superconducting properties using maximally localized Wannier functions},
	Url = {http://www.sciencedirect.com/science/article/pii/S0010465516302260},
	Year = {2016}}

@article{Noffsinger20102140,
	Author = {Jesse Noffsinger and Feliciano Giustino and Brad D. Malone and Cheol-Hwan Park and Steven G. Louie and Marvin L. Cohen},
	Doi = {http://dx.doi.org/10.1016/j.cpc.2010.08.027},
	Issn = {0010-4655},
	Journal = {Comput. Phys. Commun.},
	Number = {12},
	Pages = {2140 - 2148},
	Title = {EPW: A program for calculating the electron--phonon coupling using maximally localized Wannier functions},
	Url = {http://www.sciencedirect.com/science/article/pii/S0010465510003218},
	Volume = {181},
	Year = {2010}}

@article{Giustino2007,
	Author = {Giustino, Feliciano and Cohen, Marvin L. and Louie, Steven G.},
	Doi = {10.1103/PhysRevB.76.165108},
	Issue = {16},
	Journal = {Phys. Rev. B},
	Month = {Oct},
	Numpages = {19},
	Pages = {165108},
	Publisher = {American Physical Society},
	Title = {Electron-phonon interaction using Wannier functions},
	Url = {http://link.aps.org/doi/10.1103/PhysRevB.76.165108},
	Volume = {76},
	Year = {2007}}

@article{yang2026charge,
  title={Charge-density wave-mediated soft phonon modes suppress the temperature dependence of lattice thermal conductivity in 2H-NbSe2},
  author={Yang, Jinfeng and Tang, Zhunyun and Xu, Yongze and Wang, Xiaonan and Ouyang, Tao and Sun, Huarui},
  journal={Applied Physics Letters},
  volume={128},
  number={10},
  year={2026},
  publisher={AIP Publishing},
  doi={10.1063/5.0309984}
}

@article{Xi2015monolayerCDW,
	Author = {Xi, Xiaoxiang and Zhao, Liang and Wang, Zefang and Berger, Helmuth and Forr{\'o}, L{\'a}szl{\'o} and Shan, Jie and Mak, Kin Fai},
	Doi = {10.1038/nnano.2015.143},
	Journal = {Nat. Nanotechnol.},
	Number = {9},
	Pages = {765--769},
	Title = {Strongly enhanced charge-density-wave order in monolayer NbSe2},
	Volume = {10},
	Year = {2015}}

@article{Wilson1975CDW,
  title={Charge-density waves and superlattices in the metallic layered transition metal dichalcogenides},
  Doi = {10.1080/00018737500101391},
  author={Wilson, Jl A and Di Salvo, FJ and Mahajan, S},
  journal={Advances in physics},
  volume={24},
  number={2},
  pages={117--201},
  year={1975},
  publisher={Taylor \& Francis}
}

@article{Calandra2009NbSe2,
	Author = {Calandra, Matteo and Mazin, I. I. and Mauri, Francesco},
	Doi = {10.1103/PhysRevB.80.241108},
	Journal = {Phys. Rev. B},
	Number = {24},
	Pages = {241108},
	Title = {Effect of dimensionality on the charge-density wave in few-layer NbSe$_2$},
	Volume = {80},
	Year = {2009}}

@article{xia2026self,
  title={Self-Consistent Phonon Spectral Functions and Thermal Transport Beyond the Quasiparticle Approximation},
  author={Xia, Yi},
  journal={arXiv preprint arXiv:2607.10211},
  year={2026}
}

@article{Weber2011NbSe2,
	Author = {Weber, F. and Rosenkranz, S. and Castellan, J.-P. and Osborn, R. and Hott, R. and Heid, R. and Bohnen, K.-P. and Egami, T. and Said, A. H. and Reznik, D.},
	Doi = {10.1103/PhysRevLett.107.107403},
	Journal = {Phys. Rev. Lett.},
	Number = {10},
	Pages = {107403},
	Title = {Extended Phonon Collapse and the Origin of the Charge-Density Wave in {$2H$}-{NbSe}$_2$},
	Volume = {107},
	Year = {2011}}

@article{exists2026giant,
  title={Giant Kohn Anomaly and Chiral Phonons in the Charge-Density Wave Phase of 1H-NbSe2: Impact of Phonon Anticrossing},
  author={Exists, Susy and Mardanya, Sougata and Markiewicz, Robert and Hakioglu, Tugrul and Nieminen, Jouko and Harkoonen, Ville J and Sanga, Cem and Bansil, Arun and Chowdhury, Sugata},
  journal={ACS nano},
  volume={20},
  number={8},
  pages={6558--6571},
  year={2026},
  publisher={ACS Publications},
  doi={10.1021/acsnano.5c11110}
}

@article{Johannes2008CDW,
	Author = {Johannes, M. D. and Mazin, I. I.},
	Doi = {10.1103/PhysRevB.77.165135},
	Journal = {Phys. Rev. B},
	Number = {16},
	Pages = {165135},
	Title = {Fermi surface nesting and the origin of charge density waves in metals},
	Volume = {77},
	Year = {2008}}

@article{Ugeda2016NbSe2,
	Author = {Ugeda, Miguel M. and Bradley, Aaron J. and Zhang, Yi and Onishi, Seita and Chen, Yi and Ruan, Wei and Ojeda-Aristizabal, Claudia and Ryu, Hyejin and Edmonds, Mark T. and Tsai, Hsin-Zon and Riss, Alexander and Mo, Sung-Kwan and Lee, Dunghai and Zettl, Alex and Hussain, Zahid and Shen, Zhi-Xun and Crommie, Michael F.},
	Doi = {10.1038/nphys3527},
	Journal = {Nat. Phys.},
	Number = {1},
	Pages = {92--97},
	Title = {Characterization of collective ground states in single-layer NbSe$_2$},
	Volume = {12},
	Year = {2016}}

@article{Zheng2019NbSe2,
	Author = {Zheng, Feipeng and Feng, Ji},
	Doi = {10.1103/PhysRevB.99.161119},
	Journal = {Phys. Rev. B},
	Number = {16},
	Pages = {161119},
	Title = {Electron-phonon coupling and the coexistence of superconductivity and charge-density wave in monolayer NbSe$_2$},
	Volume = {99},
	Year = {2019}}

@article{Bianco2020NbSe2,
	Author = {Bianco, Raffaello and Monacelli, Lorenzo and Calandra, Matteo and Mauri, Francesco and Errea, Ion},
	Doi = {10.1103/PhysRevLett.125.106101},
	Journal = {Phys. Rev. Lett.},
	Number = {10},
	Pages = {106101},
	Title = {Weak Dimensionality Dependence and Dominant Role of Ionic Fluctuations in the Charge-Density-Wave Transition of NbSe$_2$},
	Volume = {125},
	Year = {2020}}

@article{rivano2026nbse2,
	Author = {Rivano, Norma and Libbi, Francesco and Tan, Chuin Wei and Cheung, Christopher T. S. and Lado, Jose L. and Mostofi, Arash A. and Kim, Philip and Lischner, Johannes and Fumega, Adolfo O. and Kozinsky, Boris and Goodwin, Zachary A. H.},
	Doi = {10.1038/s41524-026-02063-4},
	Journal = {npj Comput. Mater.},
	Title = {Exploring charge density waves in two-dimensional {NbSe}$_2$ with machine learning},
	Year = {2026}}

@article{benic2025nbse2,
	Author = {Beni{\'c}, Luka and Grasselli, Federico and Ben Mahmoud, Chiheb and Novko, Dino and Lon{\v{c}}ari{\'c}, Ivor},
	Doi = {10.1021/acs.jctc.5c00959},
	Journal = {J. Chem. Theory Comput.},
	Number = {16},
	Pages = {8130--8141},
	Title = {Machine Learning Model for Efficient Nonthermal Tuning of the Charge Density Wave in Monolayer {NbSe}$_2$},
	Volume = {21},
	Year = {2025}}

@article{Lian2018NbSe2,
	Author = {Lian, Chao-Sheng and Si, Chen and Duan, Wenhui},
	Doi = {10.1021/acs.nanolett.8b00237},
	Journal = {Nano Lett.},
	Number = {5},
	Pages = {2924--2929},
	Title = {Unveiling Charge-Density Wave, Superconductivity, and Their Competitive Nature in Two-Dimensional NbSe$_2$},
	Volume = {18},
	Year = {2018}}

@article{Manzeli2017TMD,
	Author = {Manzeli, Sajedeh and Ovchinnikov, Dmitry and Pasquier, Diego and Yazyev, Oleg V. and Kis, Andras},
	Doi = {10.1038/natrevmats.2017.33},
	Journal = {Nat. Rev. Mater.},
	Number = {8},
	Pages = {17033},
	Title = {2D transition metal dichalcogenides},
	Volume = {2},
	Year = {2017}}

@article{Borisenko2009NbSe2,
	Author = {Borisenko, S. V. and Kordyuk, A. A. and Zabolotnyy, V. B. and Inosov, D. S. and Evtushinsky, D. and B{\"u}chner, B. and Yaresko, A. N. and Varykhalov, A. and Follath, R. and Eberhardt, W. and Patthey, L. and Berger, H.},
	Doi = {10.1103/PhysRevLett.102.166402},
	Journal = {Phys. Rev. Lett.},
	Number = {16},
	Pages = {166402},
	Title = {Two Energy Gaps and Fermi-Surface ``Arcs'' in NbSe$_2$},
	Volume = {102},
	Year = {2009}}

@article{Valla2004NbSe2,
	Author = {Valla, T. and Fedorov, A. V. and Johnson, P. D. and Glans, P.-A. and McGuinness, C. and Smith, K. E. and Andrei, E. Y. and Berger, H.},
	Doi = {10.1103/PhysRevLett.92.086401},
	Journal = {Phys. Rev. Lett.},
	Number = {8},
	Pages = {086401},
	Title = {Quasiparticle Spectra, Charge-Density Waves, Superconductivity, and Electron-Phonon Coupling in {$2H$}-{NbSe}$_2$},
	Volume = {92},
	Year = {2004}}

@article{Xi2016Ising,
	Author = {Xi, Xiaoxiang and Wang, Zefang and Zhao, Wenjin and Park, Ju-Hyun and Law, Kam Tuen and Berger, Helmuth and Forr{\'o}, L{\'a}szl{\'o} and Shan, Jie and Mak, Kin Fai},
	Doi = {10.1038/nphys3538},
	Journal = {Nat. Phys.},
	Number = {2},
	Pages = {139--143},
	Title = {Ising pairing in superconducting NbSe$_2$ atomic layers},
	Volume = {12},
	Year = {2016}}

@article{Xu2018NbSe2,
	Author = {Xu, Cai-Zhi and Wang, Xiaoxiong and Chen, Peng and Fl{\"o}totto, David and Hlevyack, Joseph Andrew and Lin, Meng-Kai and Bian, Guang and Mo, Sung-Kwan and Chiang, Tai-Chang},
	Doi = {10.1103/PhysRevMaterials.2.064002},
	Journal = {Phys. Rev. Mater.},
	Number = {6},
	Pages = {064002},
	Title = {Experimental and theoretical electronic structure and symmetry effects in ultrathin NbSe$_2$ films},
	Volume = {2},
	Year = {2018}}

@article{Rossnagel2011CDW,
  title={On the origin of charge-density waves in select layered transition-metal dichalcogenides},
  author={Rossnagel, K},
  journal={Journal of Physics: Condensed Matter},
  volume={23},
  number={21},
  pages={213001},
  year={2011},
  doi={10.1088/0953-8984/23/21/213001}
}

@article{bevolo1974specific,
  title={Specific heat of 2H NbSe2},
  author={Bevolo, Albert J and Shanks, HR},
  journal={Journal of Applied Physics},
  volume={45},
  number={10},
  pages={4644--4648},
  year={1974},
  publisher={American Institute of Physics},
  Doi = {10.1063/1.1663104}
}

@article{Moncton1975NbSe2,
	Author = {Moncton, D. E. and Axe, J. D. and DiSalvo, F. J.},
	Doi = {10.1103/PhysRevLett.34.734},
	Journal = {Phys. Rev. Lett.},
	Number = {12},
	Pages = {734--737},
	Title = {Neutron Scattering Study of the Charge-Density Wave Transitions in {$2H$}-{TaSe}$_2$ and {$2H$}-{NbSe}$_2$},
	Volume = {34},
	Year = {1975}}

\end{document}